\documentclass[12pt,english]{article}
\usepackage[T1]{fontenc}
\usepackage{esint}

\makeatletter
\@ifundefined{date}{}{\date{}}

\newtheorem{theorem}{Theorem}

\newtheorem{lemma}{Lemma}

\makeatother

\usepackage{babel}
\begin{document}

\title{One-dimensional Coulomb multi-particle systems}

\author{Malyshev V.A., Zamyatin A.A.\thanks{119991, Russia,
    Moscow State Lomonosov Unoversity. Faculty of Mechanics and Mathematics,
    Leninskie Gory, Main Building 1}}

\maketitle
\tableofcontents{}
\begin{abstract}
We consider the system of particles with equal charges and nearest
neighbour Coulomb interaction on the interval. We study local properties
of this system, in particular the distribution of distances between
neighbouring charges. For zero temperature case there is sufficiently
complete picture and we give a short review. For Gibbs distribution
the situation is more difficult and we present two related results. 
\end{abstract}

\section{Introduction}

Many electric phenomena are not well understood and even might seem
mysterious. More exactly, most are not still deduced from the microscale
version of Maxwell equations on rigorous mathematical level. For example,
even in the standard direct or alternative current the electrons move
along hundred kilometers of power lines but the external accelerating
force acts only on some meters of the wire. Here what one can read
about this in ``Feynman lectures on physics'' (\cite{Feynman}.
volume 2, 16-2):

``...The force pushes the electrons along the wire. But why does
this move the galvanometer, which is so far from the force ? Because
when the electrons which feel the magnetic force try to move, they
push - by electric repulsion - the electrons a little farther down
the wire; they, in turn, repel the electrons a little farther on,
and so on for a long distance. An amazing thing. It was so amazing
to Gauss and Weber - who first built a galvanometer - that they tried
to see how far the forces in the wire would go. They strung the wire
all the way across the city...''.

This was written by the famous physicist. However, after that, this
``amazing thing'' was vastly ignored in the literature. For example,
the Drude model (that can be found in any textbook on solid state
physics, see for example \cite{AshMer}), considers free (non-interacting)
electrons and constant external accelerating force acting along all
the wire, without mention where this force (or field) comes from.

Many more questions arise. For example, why the electrons in DC moves
slowly but such stationary regime is being established almost immediately.
In \cite{Mal_analytic,Mal_selforganization} it was demonstrated rigorously
that even on the classical (not quantum) level that the stationary
and space homogeneous flow of charged particles may exist as a result
of self-organization of strongly interacting (via Coulomb repulsion)
system of electrons. This means that the field accelerating the electrons
is created by the neighboring electrons via some multiscale selforganization.

In fact, Ohm's law is formulated on the macroscale (of order one),
one-dimensional movement of $N$ electrons is described on the microscale
(of order $N^{-1}$), but the accelerating force is the corollary
of the processes on the so called submicroscale (of the order $N^{-2}).$
To show this we used only classical nonrelativistic physics - Newtonian
dynamics and Coulomb's law, but also the simplest friction mechanism,
ignoring where this friction mechanism comes from.

Besides other not solved dynamic problems like discharge, lightning,
global current, bioelectricity etc, also local and global properties
of equilibrium configurations of charged particles in external electric
fields are not at all studied (note that the mathematical part of
equilibrium statistical physics has been developed mostly on the lattice).
The equilibrium configurations can be either ground state (zero temperature)
or Gibbs states. Ground states are easier to describe and we give
short review of results in Part 1 of the paper. Study of local structure
of Gibbs configurations is at very beginning and we present in Part
2 two new results with complete proofs.

\part{Ground state configurations }

Consider systems of $N$ particles with equal charges, Coulomb interaction
and external force $F$ on an manifold. Even when there is no external
force, the problem appears to be sufficiently difficult, and was claimed
important already long ago \cite{berkenbusch}. For example, J. J.
Thomson (who discovered electron) suggested the problem of finding
such configurations on the sphere, and the answer has been known for
$N=2,3,4$ for more than 100 years, but for $N=5$ the solution was
obtained only quite recently \cite{schwartz}. 

More interesting is the case of large $N$, where the asymptotics
$N\to\infty$ is of main interest. In one-dimensional case T. J. Stieltjes
studied the problem with logarithmic interaction and found its connection
with zeros of orthogonal polynomials on the corresponding interval,
see \cite{chaitanya}, \cite{ismail}. However, the problem of finding
minimal energy configurations on two-dimensional sphere for any $N$
and power interaction (sometimes it is called the seventh problem
of S. Smale, it is also connected with the names of F. Risz and M.
Fekete) was completely solved only for quadratic interaction (see
\cite{smale}, \cite{dimitrov}, \cite{kuijlaars} and review \cite{nerattini}).
For more general compact manifolds see review \cite{korevaar}.

In this section we review recent results concerning nonzero external
force. Moreover, we consider not only global minima but even more
interesting case of local energy minima. It appears that even in the
simpler one-dimensional model with nearest neighbour interaction there
is an interesting structure of fixed points (more exactly, fixed configurations),
rich both in the number and in the charge distribution.

\section{The model}

We consider the set of configurations of $N+1$ point particles 
\[
-L\leq x_{N}<...<x_{1}<x_{0}\leq0
\]
with equal charges on the segment $[-L,0]$. Here $N$ is assumed
to be sufficiently large. We assume repulsive Coulomb interaction
of nearest neighbours, and external force $\alpha_{ext}F_{0}(x)$,
that is the potential energy is given by

\begin{equation}
U=\sum_{i=1}^{N}V(x_{i-1}-x_{i})-\sum_{i=0}^{N}\int_{-L}^{x_{i}}\alpha_{ext}F_{0}(x)dx,V(x)=\frac{\alpha_{int}}{|x|}\label{energy_U}
\end{equation}
where $\alpha_{ext},\alpha_{int}$ are positive constants. This defines
the dynamics of the system of charges, if one defines exactly what
occurs with particles $0$ and $N$ in the points $0$ and $-L$ correspondingly.
Namely, we assume completely inelastic boundary conditions. More exactly,
when particle $x_{0}(t)$ at time $t$ reaches point $0$, having
some velocity $v_{0}(t-0)\geq0$, then its velocity $v_{0}(t)$ immediately
becomes zero, and the particle itself stays at point $0$ until the
force acting on it (which varies accordingly to the motion of other
particles) becomes negative. Similarly for the particle $x_{N}(t)$
at point $-L$.

\section{Problem of many local minima}

It is evident that if $F_{0}\equiv0$, then there is only one fixed
point with 
\begin{equation}
\delta_{k}=x_{k-1}-x_{k}=\frac{L}{N},k=1,...,N\label{F_ravno_0}
\end{equation}
Thus it is the global energy minimum. More general result is the following

\begin{theorem}\label{th_uniqueness}

Assume that $F_{0}(x)$ is continuous, nonnegative and monotonic.
Then for any $N,L,\alpha_{ren}$ the fixed point exists and is unique.

\end{theorem}

However, the monotonicity assumption in this theorem is very essential.
An example of strong nonuniqueness (where the number of fixed points
is of the order of $N$) is very simple - for a function $F_{0}(x)$
with the only maximum inside the interval. Namely, on the interval
$[-1,1]$ put for $b>a>0$ 
\[
F_{0}(x)=a-2ax,x\geq0
\]
\[
F_{0}(x)=a+2bx,x\leq0
\]
Then there exists $C_{cr}>0$ such that for all sufficiently large
$N$ and $\alpha_{ren}=cN,c>C_{cr}$, one can show using similar techniques
that for any odd $N_{1}<N$ there exists fixed point such that 
\[
-1=x_{N}<...x_{N_{1}}<0<x_{N_{1}-1}<...<x_{\frac{N_{1}+1}{2}}=\frac{1}{2}<...<x_{0}<1
\]
Moreover, any such point will give local minimum of the energy.

One-dimensional case shows what can be expected in multi-dimensional
case, which is more complicated but has great interest in connection
to the static charge distribution in the atmosphere or in the live
organism.

\section{Phase transitions}

To discover phase transitions one should consider asymptotics $N\to\infty$,
with the parameters $L,F_{0}(x)$ being fixed. Then the fixed points
will depend only on the ``renormalized force'' $F=\frac{\alpha_{ext}}{\alpha_{int}}F_{o}$,
and we assume that the renormalized constant $\alpha_{ren}=\frac{\alpha_{ext}}{\alpha_{int}}$
can tend to infinity together with $N$, namely as $\alpha_{ren}=cN^{\gamma}$,
where $c,\gamma>0$. 

The necessity to consider cases when $\alpha_{ren}$ depends on $N$,
issues from concrete examples where $\alpha_{ren}\gg N$. E.g. the
linear density of electrons in some conductors, see \cite{AshMer},
is of the order $N\approx10^{9}m^{-1}$, $\alpha_{int}=\frac{e^{2}}{\epsilon_{0}}\approx10^{-28}$
and $\alpha_{_{ext}}=220\frac{volt}{meter}e=220\times10^{-19}$ (in
SI system). Thus $\alpha_{ren}$ has the order $10^{11}$. This is
close to the critical point of our model, which, as it will be shown,
is asymptotically $c_{cr}N$, that is close to $4\times10^{9}$ in
our case.

Below this section we assume for simplicity that $F_{0}>0$ is constant.
We formulate now the assertions proven in \cite{Mal_fixed_circle,Mal_fixed_MMJ,Mal_phase}.

\paragraph{Critical force}

For any $N,L$ there exists $F_{cr}=F_{cr}(N,L)$ such that for the
fixed point the following holds: $x_{N}>-L$ for $F>F_{cr}$ and $x_{N}=-L$
for $F\leq F_{cr}$. If $F=cN^{\gamma},\gamma>1,$ then for any $c>0$
we have $x_{N}\to0$. At the same time $F_{cr}\sim_{N\to\infty}c_{cr}N$,
where 
\begin{equation}
c_{cr}=\frac{4}{L^{2}}\label{c_critical}
\end{equation}

\paragraph{Multiscale phase}

The case when $\alpha_{ren}$ does not depend on $N$ was discussed
in details in \cite{Mal_fixed_circle,Mal_fixed_MMJ}, there are no
phase transitions but it is discovered that the structure of the fixed
configuration differs from (\ref{F_ravno_0}) only on the sub-micro-scale
of the order $N^{-2}$. More exactly, consider more general case when
$V(x)=|x|^{-b},b>0$. Then the following holds: if $F$ does not depend
on $N$ then for any $k=1,...,N$

\[
(x_{k-1}-x_{k})-\frac{L}{N}\sim\frac{FL^{1+b}}{1+b}N^{-b}(k-\frac{N}{2})
\]

\paragraph{Uniform density}

We define the density $\rho(x)$ so that for any subintervals $I\subset[-L,0]$
there exist the limits 
\[
\rho(I)=\int_{I}\rho(x)dx=\lim_{N\to\infty}\frac{\#\{i:x_{i}\in I\}}{N}
\]
Then if $F=o(N)$, then the density exists and is strictly uniform,
that is for all $k=1,...,N$ as $N\to\infty$ 
\begin{equation}
\max_{k}|(x_{k-1}-x_{k})-\frac{L}{N}|=o(\frac{1}{N})\label{th_2}
\end{equation}

\paragraph{Non-uniform density}

If $F=cN$ and $0<c\leq c_{cr}$, then $x_{N}=-L$ and the density
of particles exists, is nowhere zero, but is not uniform (not constant
in $x$).

\paragraph{Weak contraction}

If $F=cN$ and $c>c_{cr}$, then as $N\to\infty$ 
\begin{equation}
-L<x_{N}\to-\frac{2}{\sqrt{c}}\label{c_crit}
\end{equation}
and the density on the interval $(-\frac{2}{\sqrt{c}},0)$ is not
uniform.

\paragraph{Strong contraction}

If $F=cN^{\gamma},\gamma>1,$ then the density $\rho(x)\to\delta(x)$
in the sense of distributions. 

Both contraction cases are related to the discharge possibility, as
after disappearance of the external force, discharge can be produced
the strength of which depends on the initial concentration of charged
particles.

\part{Gibbs distribution}

We consider the set $\Omega=\Omega_{N}=\{\omega=(x_{0},...,x_{N})\},N\geq2,$
of configurations of $N+1$ points particles on the segment $[0,L]$
such that 
\begin{equation}
0=x_{0}<...<x_{N}=L\label{configuration}
\end{equation}
Introducing new variables $u_{k}=x_{k}-x_{k-1},k=1,...,N$, one sees
that $\Omega_{N}$ is an open simplex 
\[
u_{1}+...+u_{N}=L,u_{k}>0
\]
which is denoted by $S(N,L)$. 

We shall consider the probability density 
\begin{equation}
P(\omega)=Z_{N}^{-1}\exp(-\beta U(\omega))\label{gibbs}
\end{equation}
on $S(N,L)$ with respect to the Lebesgue measure $\nu$ on $S(N,L)$,
where 
\[
Z_{N}=\int_{S(N,L)}\exp(-\beta U(\omega))d\nu=\int_{0<x_{1}<...<x_{N-1}<L}\exp(-\beta U(\omega))dx_{1}...dx_{N-1}=
\]

\[
=\int_{S(N,L)}\prod_{k=1}^{N}\exp(-\beta V(u_{k}))du_{1}...du_{N}
\]
and

\[
U(\omega)=V(x_{1}-x_{0})+...+V(x_{N}-x_{N-1})=\sum_{k=1}^{N}V(u_{k})
\]
is the function on the set of configurations called the energy, and
$V$ is the function on the segment $[0,\infty]$, called potential.
Most interesting case for us is the Coulomb repulsive potential 
\[
V(u)=\frac{1}{u},u>0
\]
Equivalently one could say that we consider the sum 
\[
S_{N}=\xi_{1}+...+\xi_{N}
\]
of $N$ independent identically distributed positive random variables
$\xi_{k}$, each having density $g(u),u>0,$ further assumed to be
smooth, for simplicity. Then the conditional density of the vector
$\{\xi_{1},...,\xi_{N}\}$, under the condition $S_{N}=L$, 
\begin{equation}
P(\omega)=\frac{g(u_{1})...g(u_{N})}{\int_{S(N,L)}g(u_{1})...g(u_{N})du_{1}...du_{N}}\label{conditional}
\end{equation}
that coincides with (\ref{gibbs}), if we put 
\[
g(u)=Z_{1}^{-1}\exp(-\beta V(u)),\quad Z_{1}=\int_{0}^{L}\exp(-\beta V(u))du
\]
It is clear that conditional distributions $P(\xi_{k}<x|S_{N}=L)$
are the same for all $k$. In particular 
\[
<\xi_{k}|S_{N}=L>=\frac{L}{N}
\]
Note that in the limiting case $\beta=\infty$ the distribution is
concentrated in the unique fixed point $u_{k}=\frac{L}{N}$. 

Below we put $L=1$. Let

\[
f^{\star n}(x)=f\star f\star f\star...\star f
\]
be the $n$ times convolution of $f(x)$. Then the conditional variance
is 
\begin{equation}
d_{N}=D(\xi_{1}|S_{N}=1)=\int_{0}^{1}\left(x-\frac{1}{N}\right)^{2}\frac{g(x)g^{\star(N-1)}(1-x)}{g^{\star N}(1)}dx\label{disp}
\end{equation}

We want to note here that there exist many papers, related to the
famous Kac mean field model, where conditional independence (chaos)
of $\xi_{k}$ is proved under various conditions, see for example
\cite{carlen} and references therein. We follow here another goal
- to reveal possible multiscale local structure in the Gibbs situation,
which could resemble zero temperature case structure, discussed in
Part I.

\section{Results}

We consider the densities having the following asymptotic behaviour
as $x\to0$ 
\begin{equation}
g(x)\sim c_{0}x^{\alpha-1}e^{-\frac{\beta}{x}},\quad\alpha\in R,\;\beta\geq0\label{cond-1}
\end{equation}

\begin{theorem}\label{variance_1}

Under this condition and if $\alpha>0$ and $\beta=0$, as $N\to\infty$
\[
d_{N}\sim c_{1}N^{-2}
\]
for some constant $c_{1}=c_{1}(\alpha)>0$ depending only on $\alpha$.

\end{theorem}

\begin{theorem}\label{th2}If $\beta>0$ then for any $\alpha\in R$,
as $N\to\infty$ 
\[
d_{N}\sim c_{1}N^{-3}
\]
for some constant $c_{1}=c_{1}(\beta)>0$ depending only on $\beta$.

\end{theorem}

It is of interest to know the behaviour of the covariance for densities
with hyperexponential decrease at zero.

\section{Proofs }

\subsection{General power asymptotics}

We will prove here Theorem \ref{variance_1}. Instead of one density
$g(x)$ it is useful to consider the family of densities (such trick
has been used in some large deviations problems, see \cite{diaconis,Dembo})
\begin{equation}
h_{\lambda}(x)=e^{-\lambda x}g(x)z^{-1}(\lambda)\label{expfam}
\end{equation}
where $\lambda\geq0$ and 
\[
z(\lambda)=\int_{0}^{1}e^{-\lambda x}g(x)dx
\]
Let $\xi_{\lambda,k}$ be random variables with density $h_{\lambda}(x)$.
Put $m_{\lambda}=E\xi_{\lambda,k}$, $\sigma_{\lambda}^{2}=D\xi_{\lambda,k}$
and denote the conditional densities of $\xi_{\lambda,k}$ under the
condition that $S_{N}=1$ 
\begin{equation}
f_{\lambda}(x)=\frac{h_{\lambda}(x)h_{\lambda}^{\star(N-1)}(1-x)}{h_{\lambda}^{\star N}(1)}=\frac{g(x)g^{\star(n-1)}(1-x)}{g^{\star n}(1)},\; x\in[0,1]\label{notdepend}
\end{equation}
It is easy to check that $f_{\lambda}(x)$ in fact does not depend
on $\lambda$.

\begin{lemma}\label{l1} 
\begin{enumerate}
\item $m_{\lambda}\sim\alpha\lambda^{-1}$ and $\sigma_{\lambda}^{2}\sim\alpha\lambda^{-2}$
as $\lambda\to\infty$ 
\item there exists a unique $\lambda_{N}$, such that $m_{\lambda_{N}}=\frac{1}{N}$.
Then $\lambda_{N}\sim\alpha N$, $\sigma_{\lambda_{n}}^{2}\sim\alpha^{-1}N^{-2}$
as $N\to\infty$. 
\end{enumerate}
\end{lemma}

Proof.

1) By abelian theorem, see \cite{Feller} p. 445 Theorem 3, and the
condition $g(x)\sim c_{0}x^{\alpha-1}$ as $x\to0$ we have as $\lambda\to\infty$
\[
z(\lambda)\sim\frac{\Gamma(\alpha)c_{0}}{\lambda^{N}},\quad-z^{\prime}(\lambda)\sim\frac{\Gamma(\alpha+1)c_{0}}{\lambda^{N+1}},\quad z^{\prime\prime}(\lambda)\sim\frac{\Gamma(\alpha+2)c_{0}}{\lambda^{N+2}}
\]
It follows that 
\[
m_{\lambda}=-\frac{z^{\prime}(\lambda)}{z(\lambda)}\sim\frac{\Gamma(\alpha+1)}{\Gamma(\alpha)\lambda}=\frac{\alpha}{\lambda}
\]
and 
\[
\sigma_{\lambda}^{2}=\frac{c^{\prime\prime}(\lambda)}{c(\lambda)}-\left(\frac{c^{\prime}(\lambda)}{c(\lambda)}\right)^{2}\sim\left(\frac{\Gamma(\alpha+2)}{\Gamma(\alpha)}-\left(\frac{\Gamma(\alpha+1)}{\Gamma(\alpha)}\right)^{2}\right)\lambda^{-2}=\frac{\alpha^{2}}{\lambda^{2}}
\]
as $\lambda\to\infty.$

2) The function $m_{\lambda}$ is monotonically decreasing in $\lambda.$
Thus for any $N$ there exists $\lambda_{N}$ such that 
\[
m_{\lambda_{N}}=\frac{1}{N}
\]
From 1) it follows that $\lambda_{N}\sim\alpha N$ and $\sigma_{\lambda_{N}}^{2}\sim\alpha^{-1}N^{-2}$
as $N\to\infty$.

Let $\phi_{\lambda}(t)$ be the characteristic function of $\xi_{\lambda}$.

\begin{lemma}\label{lemexpfam}The family of densities (\ref{expfam}) has the
following properties: 
\begin{enumerate}
\item the normalized moment $a_{\lambda}=\frac{E|\xi_{\lambda}-m_{\lambda}|^{4}}{\sigma_{\lambda}^{4}}$
is bounded uniformly in $\lambda>0$. 
\item for any $\delta>0$ there exists $\lambda_{0}>0$ such that $\sup_{\lambda>\lambda_{0}}\sup_{t>\delta}\phi_{\lambda}(t/\sigma_{\lambda})<1$. 
\item for some $q\geq1$ 
\[
\int_{-\infty}^{\infty}|\phi_{\lambda}(t)|^{q}dt=O(\lambda^{q\alpha}).
\]

\end{enumerate}
\end{lemma}

Proof.

1) It is similar to the proof of 1) in lemma \ref{l1}.

2) Put 
\[
f(t,\lambda)=\phi_{\lambda}(t\sigma_{\lambda}^{-1})=z^{-1}(\lambda)\int_{0}^{1}e^{it\sigma_{\lambda}^{-1}x}e^{-\lambda x}g(x)dx
\]
Let us show that for some $\delta>0$ 
\begin{equation}
|f(t,\lambda)|\leq\left|\frac{1}{1-it\alpha^{-1/2}}\right|^{\alpha}+O\left(e^{-\delta\sqrt{\lambda}}\right)\label{point}
\end{equation}
as $\lambda\to\infty$. For this can write $f(t,\lambda)$ as 
\[
f(t,\lambda)=J_{1}(\lambda)+J_{2}(\lambda)
\]
where 
\[
J_{1}(\lambda)=z^{-1}(\lambda)\int_{0}^{\lambda^{-1/2}}e^{it\sigma_{\lambda}^{-1}x}e^{-\lambda x}g(x)dx
\]
\[
J_{2}(\lambda)=z^{-1}(\lambda)\int_{\lambda^{-1/2}}^{1}e^{it\sigma_{\lambda}^{-1}x}e^{-\lambda x}g(x)dx
\]
Taking into account that 
\[
z(\lambda)\sim c_{0}\Gamma(\alpha)\lambda^{-\alpha},\;\sigma_{\lambda}^{2}\sim\alpha\lambda^{-2},\;\lambda\to\infty
\]
and condition $g(x)\sim c_{0}x^{\alpha-1}$ as $x\to0$ we have that
as $\lambda\to\infty$ 
\[
\left|J_{1}(\lambda)\right|\sim(\Gamma(\alpha))^{-1}\lambda^{\alpha}\left|\int_{0}^{\lambda^{-1/2}}e^{it\alpha^{-1/2}\lambda x}e^{-\lambda x}x^{\alpha-1}dx\right|
\]
Putting $y=\lambda x$ in the last integral we get 
\[
(\Gamma(\alpha))^{-1}\lambda^{\alpha}\int_{0}^{\lambda^{-1/2}}e^{it\alpha^{-1/2}\lambda x}e^{-\lambda x}x^{\alpha-1}dx=(\Gamma(\alpha))^{-1}\int_{0}^{\lambda^{1/2}}e^{it\alpha^{-1/2}y}e^{-y}y^{\alpha-1}dy
\]
As 
\begin{eqnarray*}
(\Gamma(\alpha))^{-1}\int_{0}^{\lambda^{1/2}}e^{it\alpha^{-1/2}y}e^{-y}y^{\alpha-1}dy & = & (\Gamma(\alpha))^{-1}\int_{0}^{\infty}e^{it\alpha^{-1/2}y}e^{-y}y^{\alpha-1}dy\\
 & - & (\Gamma(\alpha))^{-1}\int_{\lambda^{1/2}}^{\infty}e^{it\alpha^{-1/2}y}e^{-y}y^{\alpha-1}dy
\end{eqnarray*}
and 
\[
(\Gamma(\alpha))^{-1}\int_{0}^{\infty}e^{it\alpha^{-1/2}y}e^{-y}y^{\alpha-1}dy=\left(\frac{1}{1-it\alpha^{-1/2}}\right)^{\alpha}
\]

\[
(\Gamma(\alpha))^{-1}\left|\int_{\lambda^{1/2}}^{\infty}e^{it\alpha^{-1/2}y}e^{-y}y^{\alpha-1}dy\right|=O\left(e^{-\delta\sqrt{\lambda}}\right)
\]
for some $\delta>0$, then 
\[
\left|J_{1}(\lambda)\right|\leq\left|\frac{1}{1-it\alpha^{-1/2}}\right|^{\alpha}+O\left(e^{-\delta\sqrt{\lambda}}\right)
\]
For $J_{2}(\lambda)$ we get the estimate 
\[
\left|J_{2}(\lambda)\right|=c^{-1}(\lambda)\left|\int_{\lambda^{-1/2}}^{1}e^{it\sigma_{\lambda}^{-1}x}e^{-\lambda x}g(x)dx\right|\leq C\lambda^{\alpha}e^{-\sqrt{\lambda}}=O\left(e^{-\delta\sqrt{\lambda}}\right)
\]
From these estimates (\ref{point}) follows.

3) Note that always 
\[
\int_{0}^{1}|h_{\lambda}(x)|^{p}dx<\infty
\]
for some $p>1.$ Without loss of generality one can assume that $1<p\leq2.$
By Hausdorff-Young inequality 
\[
\left(\int_{-\infty}^{\infty}|\phi_{\lambda}(t)|^{q}dt\right)^{\frac{1}{q}}\leq\left(\frac{1}{2\pi}\int_{0}^{1}|h_{\lambda}(x)|^{p}dx\right)^{\frac{1}{p}}
\]
where$1<p\leq2$ and $\frac{1}{p}+\frac{1}{q}=1$. As 
\[
\left(\frac{1}{2\pi}\int_{0}^{1}|h_{\lambda}(x)|^{p}dt\right)^{\frac{1}{p}}=z^{-1}(\lambda)\left(\frac{1}{2\pi}\int_{0}^{1}|e^{-\lambda x}g(x)|^{p}dt\right)^{\frac{1}{p}}\leq z^{-1}(\lambda)\left(\frac{1}{2\pi}\int_{0}^{1}|g(x)|^{p}dt\right)^{\frac{1}{p}}
\]
and $z(\lambda)\sim C\lambda^{-\alpha}$, then 
\[
\int_{-\infty}^{\infty}|\phi_{\lambda}(t)|^{q}dt=O(\lambda^{q\alpha})
\]

\begin{lemma}\label{cpt}Assume conditions 1--3 of lemma \ref{lemexpfam}
and that $\lambda_{N}$ are defined by the condition $m_{\lambda_{N}}=\frac{1}{N}$.
Then 
\[
h_{\lambda_{N}}^{\star N}(x)=\frac{1}{\sigma_{\lambda_{N}}\sqrt{2\pi N}}\left(\exp\left(-\frac{(x-1)^{2}}{2\sigma_{\lambda_{N}}^{2}N}\right)+o(1)\right),
\]
where $o(1)$ tends to $0$ uniformly in $x\geq0$.

\end{lemma}

Proof. We change a bit the standard proof of local limit theorem.
Let $p_{N}(z)$ be the density of the standard deviation 
\[
\frac{S_{N,\lambda_{N}}-Nm_{\lambda_{N}}}{\sqrt{N}\sigma_{\lambda_{N}}}
\]
where $S_{N,\lambda}=\xi_{1,\lambda}+\dots+\xi_{N,\lambda}$ is the
sum independent random variables having density (\ref{expfam}). Let
$q(z)$ be the standard Gaussian density. The inverse Fourier transform
gives 
\[
\sup_{z\in R}\left|p_{N}(z)-q(z)\right|\leq\frac{1}{2\pi}\int_{-\infty}^{\infty}\left|\psi_{\lambda_{N}}^{N}\left(\frac{t}{\sqrt{N}\sigma_{\lambda_{N}}}\right)-\frac{1}{\sqrt{2\pi}}e^{-\frac{t^{2}}{2}}\right|dt
\]
where $\psi_{\lambda}(t)=\phi_{\lambda}(t)e^{-itm_{\lambda}}$. Denote
\[
I_{1}=\frac{1}{2\pi}\int_{|t|\leq\sqrt{N}a_{\lambda_{N}}^{-1}}\left|\psi_{\lambda_{N}}^{N}\left(\frac{t}{\sqrt{N}\sigma_{\lambda_{N}}}\right)-\frac{1}{\sqrt{2\pi}}e^{-\frac{t^{2}}{2}}\right|dt
\]
\[
I_{2}=\frac{1}{2\pi}\int_{|t|>\sqrt{N}a_{\lambda_{N}}^{-1}}\left|\psi_{\lambda_{N}}\left(\frac{t}{\sqrt{N}\sigma_{\lambda_{N}}}\right)\right|^{N}dt
\]
\[
I_{3}=\frac{1}{(2\pi)^{3/2}}\int_{|t|>\sqrt{N}a_{\lambda_{N}}^{-1}}e^{-\frac{t^{2}}{2}}dt
\]
where $a_{\lambda}=\frac{E|\xi_{\lambda}-m_{\lambda}|^{4}}{\sigma_{\lambda}^{4}}$
is bounded uniformly in $\lambda$ (lemma \ref{lemexpfam}, part 1).
Then 
\[
\sup_{z\in R}\left|p_{N}(z)-q(z)\right|\leq I_{1}+I_{2}+I_{3}
\]
For $I_{1}$ we have the estimate (see \cite{Petrov} p. 109, lemma
1) 
\[
I_{1}\leq\frac{a_{\lambda_{N}}}{2\pi\sqrt{N}}\int_{|t|\leq\sqrt{N}a_{\lambda_{N}}^{-1}}|t|^{3}e^{-t^{3}/3}dt\leq\frac{Ca_{\lambda_{N}}}{\sqrt{N}}
\]
For $I_{2}$ we have by parts 2 and 3 of lemma \ref{lemexpfam}, 
\[
I_{2}=\frac{1}{2\pi}\int_{|t|>\sqrt{N}a_{\lambda_{N}}^{-1}}\left|\phi_{\lambda_{N}}\left(\frac{t}{\sqrt{N}\sigma_{\lambda_{N}}}\right)\right|^{N}dt\leq N^{-\frac{1}{2}}\gamma^{N-q}\int_{-\infty}^{\infty}\left|\phi_{\lambda_{N}}\left(t\right)\right|^{q}dt\leq CN^{-\frac{1}{2}}\gamma^{N-q}N^{q\alpha}
\]
for $N$ sufficiently large, where $\gamma<1$ and $q>1.$

The estimate for $I_{3}$ is trivial. Thus, 
\[
p_{N}(x)=q(x)+O(N^{-1/2})
\]
where $O(N^{-1/2})$ does not depend on $x\in R$. Lemma follows as
\[
h_{\lambda_{N}}^{\star N}(x)=\frac{1}{\sqrt{N}\sigma_{\lambda_{N}}}p_{N}\left(\frac{x-1}{\sqrt{N}\sigma_{\lambda_{N}}}\right)
\]
The lemma is proved.

\begin{lemma}\label{disperlemma}Assume conditions 1--3 of lemma
\ref{lemexpfam}. Then the conditional variance
\[
d_{N}=D_{N}+o(\sigma_{\lambda_{N}}^{2}),
\]
where
\begin{equation}
D_{N}=\int_{0}^{1}\left(x-\frac{1}{N}\right)^{2}h_{\lambda_{N}}(x)\exp\left(-\frac{\left(x-\frac{1}{N}\right){}^{2}}{2\sigma_{\lambda_{N}}^{2}(N-1)}\right)dx\label{dn}
\end{equation}

\end{lemma}

Proof. By lemma \ref{cpt} we have 
\[
h_{\lambda_{N}}^{\star(N-1)}(1-x)=\frac{1}{\sigma_{\lambda_{N}}\sqrt{2\pi(N-1)}}\left(\exp\left(-\frac{\left(x-\frac{1}{N}\right){}^{2}}{2\sigma_{\lambda_{N}}^{2}(N-1)}\right)+o(1)\right)
\]

\[
h_{\lambda_{N}}^{\star N}(1)=\frac{1}{\sigma_{\lambda_{N}}\sqrt{2\pi N}}\left(1+o(1)\right)
\]
where $m_{\lambda_{N}}=\frac{1}{N}$. The division gives 
\begin{equation}
\frac{h_{\lambda_{N}}^{\star(N-1)}(1-x)}{h_{\lambda_{N}}^{\star N}(1)}=\sqrt{\frac{N}{N-1}}\exp\left(-\frac{\left(x-\frac{1}{N}\right){}^{2}}{2\sigma_{\lambda_{N}}^{2}(N-1)}\right)+o(1)\label{ratio}
\end{equation}
where the term $o(1)$ tends to $0$ uniformly in $x\in[0,1].$ By
(\ref{notdepend}) the conditional variance is 
\[
d_{N}=\int_{0}^{1}\left(x-\frac{1}{N}\right)^{2}\frac{h_{\lambda_{N}}(x)h_{\lambda_{N}}^{\star(N-1)}(1-x)}{h_{\lambda_{N}}^{\star N}(1)}dx
\]
Substituting (\ref{ratio}) to this expression we get 
\[
d_{N}=\int_{0}^{1}\left(x-\frac{1}{N}\right)^{2}h_{\lambda_{N}}(x)\exp\left(-\frac{\left(x-\frac{1}{N}\right){}^{2}}{2\sigma_{\lambda_{N}}^{2}(N-1)}\right)dx+o(\sigma_{\lambda_{N}}^{2})
\]
The lemma is proved.

By lemma \ref{l1} $\sigma_{\lambda_{N}}^{2}\sim\alpha^{-1}N^{-2}$
as $N\to\infty$. Thus 
\[
d_{N}=D_{N}+o(N^{-2})
\]
where 
\[
D_{N}=\int_{0}^{1}\left(x-\frac{1}{N}\right)^{2}h_{\lambda_{N}}(x)\exp\left(-\frac{\left(x-\frac{1}{N}\right){}^{2}}{2\sigma_{\lambda_{N}}^{2}(N-1)}\right)dx
\]

\begin{lemma}As $N\to\infty$: 
\[
D_{N}\sim c_{1}N^{-2}
\]

\end{lemma}

Proof. By definition (\ref{expfam}) and lemma \ref{l1} as $N\to\infty$
\[
D_{N}\sim\frac{(\alpha N)^{\alpha}}{c_{0}\Gamma(\alpha)}\int_{0}^{1}\left(x-\frac{1}{N}\right)^{2}e^{-\alpha Nx}g(x)\exp\left(-\frac{\alpha N\left(x-\frac{1}{N}\right){}^{2}}{2}\right)dx
\]
Because of $g(x)\sim c_{0}x^{\alpha-1}$ as $x\to0$ we have 
\[
D_{N}\sim\frac{(\alpha N)^{\alpha}e^{-\alpha}}{\Gamma(\alpha)}\int_{0}^{N^{-1}}\left(x-\frac{1}{N}\right)^{2}x^{\alpha-1}dx
\]
Changing variable $y=Nx$ we have 
\[
\frac{(\alpha N)^{\alpha}e^{-\alpha}}{\Gamma(\alpha)}\int_{0}^{N^{-1}}\left(x-\frac{1}{N}\right)^{2}x^{\alpha-1}dx=\frac{\alpha{}^{\alpha}e^{-\alpha}}{\Gamma(\alpha)N^{2}}\int_{0}^{1}\left(1-y\right)^{2}y^{\alpha-1}dy
\]
which gives the lemma and the theorem.

\subsection{Coulomb case}

Again we introduce the exponential family of densities 
\begin{equation}
h_{\lambda}(x)=e^{-\lambda x}g(x)Z^{-1}(\lambda)\;\lambda\geq0\label{expfam-1}
\end{equation}
where the function $g(x)$ satisfies condition (\ref{cond-1}) with
$\beta>0$ and 
\[
Z(\lambda)=\int_{0}^{1}e^{-\lambda x}g(x)dx
\]

We will use modified Bessel function of the second kind defined as
follows: 
\begin{equation}
K_{\alpha}(z)=\frac{1}{2}\int_{0}^{\infty}x^{\alpha-1}e^{-\frac{z}{2}\left(x+\frac{1}{x}\right)}dx\label{bessel}
\end{equation}
where $\alpha\in R$ and $z>0$. For $K_{\alpha}(z)$ we know the
asymptotic expansion as $z\to\infty$ 
\begin{equation}
K_{\alpha}(z)\sim\sqrt{\frac{\pi}{2}}\frac{e^{-z}}{\sqrt{z}}\left(1+\frac{4\alpha^{2}-1}{8z}+...\right)\label{asym}
\end{equation}

Let again $\xi_{\lambda}$ be a random variable with the density $h_{\lambda}(x)$
and put $m_{\lambda}=E\xi_{\lambda}$ and $\sigma_{\lambda}^{2}=D\xi_{\lambda}.$

\begin{lemma}\label{l1-1} 
\begin{enumerate}
\item as $\lambda\to\infty$
\begin{equation}
Z(\lambda)\sim c_{0}\sqrt{\frac{\pi}{2}}\left(\frac{\beta}{\lambda}\right)^{\alpha/2}\frac{e^{-2\sqrt{\lambda\beta}}}{\left(\lambda\beta\right)^{1/4}}\label{asym1}
\end{equation}

\item as $\lambda\to\infty$ 
\[
m_{\lambda}\sim\sqrt{\frac{\beta}{\lambda}},\quad\sigma_{\lambda}^{2}\sim\frac{1}{2\beta^{1/2}\lambda^{3/2}}
\]

\item there exists a unique $\lambda_{N}$ such that $m_{\lambda_{N}}=\frac{1}{N}$
such that $\lambda_{N}\sim\beta N^{2}$ and $\sigma_{\lambda_{N}}^{2}\sim2^{-1}\beta^{-2}N^{-3}$
as $N\to\infty$. 
\end{enumerate}
\end{lemma}

Proof. 1) We can write 
\[
Z(\lambda)=I_{1}(\lambda)+I_{2}(\lambda)
\]
where 
\[
I_{1}(\lambda)=\int_{0}^{\lambda^{-\epsilon}}e^{-\lambda x}g(x)dx,\quad I_{2}(\lambda)=\int_{\lambda^{-\epsilon}}^{1}e^{-\lambda x}g(x)dx
\]
and $\epsilon>0$ is small enough. By (\ref{cond-1}) 
\[
I_{1}(\lambda)\sim c_{0}I_{1}^{\prime}(\lambda)=c_{0}\int_{0}^{\lambda^{-\epsilon}}e^{-\lambda x-\frac{\beta}{x}}x^{\alpha-1}dx
\]
as $\lambda\to\infty$. One can write 
\[
I_{1}^{\prime}(\lambda)=Z_{1}(\lambda)-Z_{1}^{\prime}(\lambda)
\]
where
\[
Z_{1}(\lambda)=\int_{0}^{\infty}e^{-\lambda x-\frac{\beta}{x}}x^{\alpha-1}dx,\quad Z_{1}^{\prime}(\lambda)=\int_{\lambda^{-\epsilon}}^{\infty}e^{-\lambda x-\frac{\beta}{x}}x^{\alpha-1}dx
\]
Find asymptotics of $Z_{1}(\lambda)$ as $\lambda\to\infty$. Changing
variable $y=\beta^{-1}x$ gives 
\[
Z_{1}(\lambda)=\beta^{\alpha}\int_{0}^{\infty}e^{-\lambda\beta y-\frac{1}{y}}y^{\alpha-1}dy
\]
and changing variable $t=\frac{2}{z}x$ in (\ref{bessel}) gives 
\[
K_{\alpha}(z)=2^{-\alpha-1}z^{\alpha}\int_{0}^{\infty}t^{\alpha-1}e^{-\frac{z^{2}}{4}t-\frac{1}{t}}dt
\]

Put $z=2\sqrt{\lambda\beta}$. Then 
\begin{equation}
Z_{1}(\lambda)=\frac{2\beta^{\alpha/2}K_{\alpha}(2\sqrt{\lambda\beta})}{\lambda^{\alpha/2}}\label{besf}
\end{equation}
and using (\ref{asym}) we get
\[
Z_{1}(\lambda)\sim\sqrt{\frac{\pi}{2}}\left(\frac{\beta}{\lambda}\right)^{\alpha/2}\frac{e^{-2\sqrt{\lambda\beta}}}{\left(\lambda\beta\right)^{1/4}},\quad\lambda\to\infty
\]
Taking into account $I_{2}(\lambda)=O(e^{-\lambda^{1-\epsilon}})$
and $Z_{1}^{\prime}(\lambda)=O(e^{-\lambda^{1-\epsilon}})$ for some
small enough $\epsilon>0$ we come to (\ref{asym1}).

2)Mathematical expectation is
\[
m_{\lambda}=Z^{-1}(\lambda)\int_{0}^{1}xe^{-\lambda x}g(x)dx
\]
By part 1 of this lemma we have
\[
\int_{0}^{1}xe^{-\lambda x}g(x)dx\sim\sqrt{\frac{\pi}{2}}\left(\frac{\beta}{\lambda}\right)^{(\alpha+1)/2}\frac{e^{-2\sqrt{\lambda\beta}}}{\left(\lambda\beta\right)^{1/4}},\quad\lambda\to\infty
\]
So 
\begin{equation}
m_{\lambda}\sim\sqrt{\frac{\beta}{\lambda}},\;\lambda\to\infty\label{mlambda}
\end{equation}

Covariance is equal to
\[
\sigma_{\lambda}^{2}=Z^{-1}(\lambda)\int_{0}^{1}x^{2}e^{-\lambda x}g(x)dx-Z^{-2}(\lambda)\left(\int_{0}^{1}xe^{-\lambda x}g(x)dx\right)^{2}
\]
It follows from part 1 that as $\lambda\to\infty$ 
\[
\sigma_{\lambda}^{2}\sim\frac{\beta}{\lambda}\left(\frac{K_{\alpha+2}(2\sqrt{\lambda\beta})}{K_{\alpha}(2\sqrt{\lambda\beta})}-\left(\frac{K_{\alpha+1}(2\sqrt{\lambda\beta})}{K_{\alpha}(2\sqrt{\lambda\beta})}\right)^{2}\right)
\]
Using asymptotic expansion (\ref{asym}) we get 
\begin{equation}
\sigma_{\lambda}^{2}\sim\frac{1}{2\beta^{1/2}\lambda^{3/2}}\label{sigmal}
\end{equation}
as $\lambda\to\infty$

3) It follows from continuity of $m_{\lambda}$ as function of $\lambda$
and (\ref{mlambda}) that there exists $\lambda_{N}$ such that $m_{\lambda_{N}}=\frac{1}{N}$
and then $\lambda_{N}\sim\beta N^{2}$ as $N\to\infty$. By (\ref{sigmal})
\[
\sigma_{\lambda_{N}}^{2}\sim\frac{1}{2\beta^{2}N^{3}},\; N\to\infty.
\]
Lemma is proved.

\begin{lemma}\label{lemexpfam-1}The exponential family (\ref{expfam-1})
has the following properties: 
\begin{enumerate}
\item the normalized moment $a_{\lambda}=\frac{E|\xi_{\lambda}-m_{\lambda}|^{4}}{\sigma_{\lambda}^{4}}$
is bounded uniformly in $\lambda>0$. 
\item for any $\delta>0$ there exists $\lambda_{0}>0$ such that $\sup_{\lambda>\lambda_{0}}\sup_{t>\delta}\phi_{\lambda}(t/\sigma_{\lambda})<1$. 
\item for some $q\geq1$ as $\lambda\to\infty$ 
\[
\int_{-\infty}^{\infty}|\phi_{\lambda}(t)|^{q}dt=O(\lambda^{q\alpha}).
\]

\end{enumerate}
\end{lemma}

Proof. Similar to lemma \ref{lemexpfam}. 

Using lemmas \ref{cpt}, \ref{disperlemma}, \ref{l1-1} we find 
\[
d_{N}=D_{N}+o(N^{-3})
\]
where 
\[
D_{N}=\int_{0}^{1}\left(x-\frac{1}{N}\right)^{2}h_{\lambda_{N}}(x)\exp\left(-\frac{\left(x-\frac{1}{N}\right){}^{2}}{2\sigma_{\lambda_{N}}^{2}(N-1)}\right)dx
\]
By lemma \ref{l1-1} $\lambda_{N}\sim\beta N^{2}$ and $\sigma_{\lambda_{N}}^{2}\sim2^{-1}\beta^{-2}N^{-3}$
as $N\to\infty$, then
\begin{equation}
D_{N}\sim Z^{-1}(\lambda_{N})\int_{0}^{1}\left(x-\frac{1}{N}\right)^{2}e^{-\beta N^{2}x}g(x)x^{\alpha-1}\exp\left(-\beta^{2}N{}^{2}\left(x-\frac{1}{N}\right)^{2}\right)dx\label{asym2}
\end{equation}
We split the integral in (\ref{asym2}) into two integrals
\begin{eqnarray*}
D_{N}^{(1)} & = & \int_{0}^{N^{-1/2}}\left(x-\frac{1}{N}\right)^{2}e^{-\beta N^{2}x}g(x)x^{\alpha-1}\exp\left(-\beta^{2}N{}^{2}\left(x-\frac{1}{N}\right)^{2}\right)dx\\
D_{N}^{(2)} & = & \int_{N^{-1/2}}^{1}\left(x-\frac{1}{N}\right)^{2}e^{-\beta N^{2}x}g(x)x^{\alpha-1}\exp\left(-\beta^{2}N{}^{2}\left(x-\frac{1}{N}\right)^{2}\right)dx
\end{eqnarray*}
By condition (\ref{cond-1}) 
\[
Z^{-1}(\lambda_{N})D_{N}^{(1)}\sim Z^{-1}(\lambda_{N})c_{0}\int_{0}^{N^{-1/2}}\left(x-\frac{1}{N}\right)^{2}e^{-\beta\left(\frac{1}{x}+N^{2}x\right)}x^{\alpha-1}\exp\left(-\beta^{2}N{}^{2}\left(x-\frac{1}{N}\right)^{2}\right)dx
\]
To find the asymptotics of $D_{N}^{(1)}$ we use Laplace's method.
Consider the function $s(x)=\beta\left(\frac{1}{x}+N^{2}x\right)$.
Its derivative 
\[
s^{\prime}(x)=\beta\left(-\frac{1}{x^{2}}+N^{2}\right)
\]
equals $0$ at the point $N^{-1}$. The second derivative 
\[
s^{\prime\prime}(x)=\frac{2\beta}{x^{3}}
\]
The function $s(x)$ can be expanded at the neighborhood of $N^{-1}$
using Taylor's formula: 
\[
s(x)=2\beta N+\beta N^{3}\left(x-\frac{1}{N}\right)^{2}+O\left(\left(x-\frac{1}{N}\right)^{3}\right)
\]
By (\ref{asym1}) we have 
\[
Z^{-1}(\lambda_{N})\sim c_{0}^{-1}\sqrt{\frac{2\beta}{\pi}}e^{2\beta N}N^{\alpha+\frac{1}{2}}
\]
and so we get as $N\to\infty$ 
\[
Z^{-1}(\lambda_{N})D_{N}^{(1)}\sim c_{0}^{-1}\sqrt{\frac{2\beta}{\pi}}e^{2\beta N}N^{\alpha+\frac{1}{2}}c_{0}\int_{0}^{1}\left(x-\frac{1}{N}\right)^{2}N^{-\alpha+1}e^{-2\beta N-\beta N^{3}\left(x-\frac{1}{N}\right)^{2}}dx
\]
After cancellations 
\[
Z^{-1}(\lambda_{N})D_{N}^{(1)}\sim\sqrt{\frac{2\beta}{\pi}}N^{\frac{3}{2}}\int_{0}^{1}\left(x-\frac{1}{N}\right)^{2}e^{-N^{3}\left(x-\frac{1}{N}\right)^{2}}dx\sim\sqrt{\frac{\beta}{2}}N^{-3}
\]
as $N\to\infty$. For the second integral $D_{N}^{(2)}$ we have as
$N\to\infty$ 
\[
Z^{-1}(\lambda_{N})D_{N}^{(2)}=O(e^{-\beta\sqrt{N}})
\]
So
\[
D_{N}\sim\sqrt{\frac{\beta}{2}}N^{-3},\: N\to\infty
\]
Theorem is proved.

\end{document}